\documentclass[prl,superscriptaddress, amsmath, amssymb,floatfix]{revtex4-1}
\usepackage{graphicx,nicefrac}
\usepackage[thinspace,thinqspace,amssymb]{SIunits}
\usepackage{xcolor,colortbl}
\usepackage{amsthm}
\usepackage{hyperref}

\date{\today{}}



\begin{document}

\title{Interface Engineering to Create a Strong Spin Filter Contact to Silicon}

\author{C. Caspers}
	\affiliation{Peter Gr\"unberg Institut (PGI-6), Forschungszentrum J\"ulich, 52425 J\"ulich, Germany}
	\affiliation{JARA J\"ulich-Aachen Research Alliance, Forschungszentrum J\"ulich, 52425 J\"ulich, Germany}
\author{A. Gloskovskii}
	\affiliation{DESY Photon Science, Deutsches Elektronen-Synchrotron, 22603 Hamburg, Germany }
\author{M. Gorgoi}
	\affiliation{Helmholtz-Zentrum f\"ur Materialien und Energie, BESSY II, Berlin, Germany}
\author{C. Besson}
	\affiliation{Peter Gr\"unberg Institut (PGI-6), Forschungszentrum J\"ulich, 52425 J\"ulich, Germany}
\author{M. Luysberg}
	\affiliation{Peter Gr\"unberg Institut (PGI-5) und Ernst Ruska-Centrum f\"ur Mikroskopie und Spektroskopie mit Elektronen, Forschungszentrum J\"ulich, 52425 J\"ulich, Germany}
\author{K. Z. Rushchanskii}
	\affiliation{Peter Gr\"unberg Institut (PGI-1), Forschungszentrum J\"ulich, 52425 J\"ulich, Germany}
	\affiliation{JARA J\"ulich-Aachen Research Alliance, Forschungszentrum J\"ulich, 52425 J\"ulich, Germany}
\author{M. Le\v{z}ai\'{c}}
	\affiliation{Peter Gr\"unberg Institut (PGI-1), Forschungszentrum J\"ulich, 52425 J\"ulich, Germany}
	\affiliation{JARA J\"ulich-Aachen Research Alliance, Forschungszentrum J\"ulich, 52425 J\"ulich, Germany}
\author{C. S. Fadley}
	\affiliation{Department of Physics, University of California, Davis, California, USA}
	\affiliation{Materials Sciences Division, Lawrence Berkeley National Laboratory, Berkeley, California, USA}
\author{W. Drube}
	\affiliation{DESY Photon Science, Deutsches Elektronen-Synchrotron, 22603 Hamburg, Germany }
\author{M. M\"uller}
	\affiliation{Peter Gr\"unberg Institut (PGI-6), Forschungszentrum J\"ulich, 52425 J\"ulich, Germany}
	\affiliation{JARA J\"ulich-Aachen Research Alliance, Forschungszentrum J\"ulich, 52425 J\"ulich, Germany}
	\affiliation{Fakult\"at f\"ur Physik, Universit\"at Duisburg-Essen, 47048 Duisburg, Germany}




\begin{abstract}
	Integrating epitaxial and ferromagnetic Europium Oxide (EuO) directly on silicon is a perfect route to enrich silicon nanotechnology with spin filter functionality. 
	To date, the inherent chemical reactivity between EuO and Si has prevented a heteroepitaxial integration without significant contaminations of the interface with Eu silicides and Si oxides.
	We present a solution to this long-standing problem by applying two complementary passivation techniques for the reactive EuO/Si interface: 
	(\emph{i}) an \emph{in situ} hydrogen-Si(001) passivation and (\emph{ii}) the application of oxygen-protective Eu monolayers --- without using any additional buffer layers.
	By careful chemical depth profiling of the oxide-semiconductor interface via hard x-ray photoemission spectroscopy, we show how to systematically minimize both Eu silicide and Si oxide formation to the sub-monolayer regime --- and how to ultimately interface-engineer chemically clean, heteroepitaxial and ferromagnetic EuO/Si (001) in order to create a strong spin filter contact to silicon.
\end{abstract}

\maketitle

	\section{Introduction}
		\label{sec:introduction}
		Interfacing magnetic oxides (MO) directly with silicon can fuel charge-based electronics with additional spin functionality.\cite{Jansen2012,Schmehl2007,Bader2010,Mueller2011} 
		Essential for the efficient spin injection and detection in any silicon spintronic device are well-defined chemical and structural properties of the MO/Si transport interface.\cite{Miao2011} 
		Silicon, however, notoriously challenges experimentalists by its high chemical reactivity, in particular at elevated temperatures, which in turn are mandatory for the high-quality synthesis of ultrathin magnetic oxides.
		So far, these contrary terms have hampered the heteroepitaxial integration of spin-functional magnetic oxides in silicon spintronics devices.

		According to theoretical predictions,\cite{Hubbard1996} the magnetic oxide Europium Oxide (EuO) is the only binary MO thermodynamically stable in direct contact with silicon. 
		EuO unites the rare combination of ferromagnetic order ($T_{\mathrm{C}} = \unit{69.3}{\kelvin}$)\cite{Kornblit1975} and insulating properties.\cite{Altendorf2011}
		In the few nanometer regime, EuO acts as a highly efficient spin-selective tunnel barrier\cite{Steeneken2002} due to its exchange-split conduction band of \unit{0.6}{\electronvolt}.\cite{Steeneken2002,Miao2011}
		Moreover, its insulating band gap of $E_{\mathrm{g}} = \unit{1.12}{\electronvolt}$\cite{Ghosh2004} matches that of silicon very well ($E_{\mathrm{g}}=\unit{1.11}{\electronvolt}$), thus overcoming the well-known conductivity mismatch problem.\cite{Schmidt2005} 
		Therefore, ultrathin EuO films are perfectly suited as efficient spin filter contacts directly on clean silicon wafers.\cite{Schmehl2007,Mueller2009,Miao2009,Mueller2011}

		\begin{figure*}
			\centering
			\includegraphics[width=\linewidth]{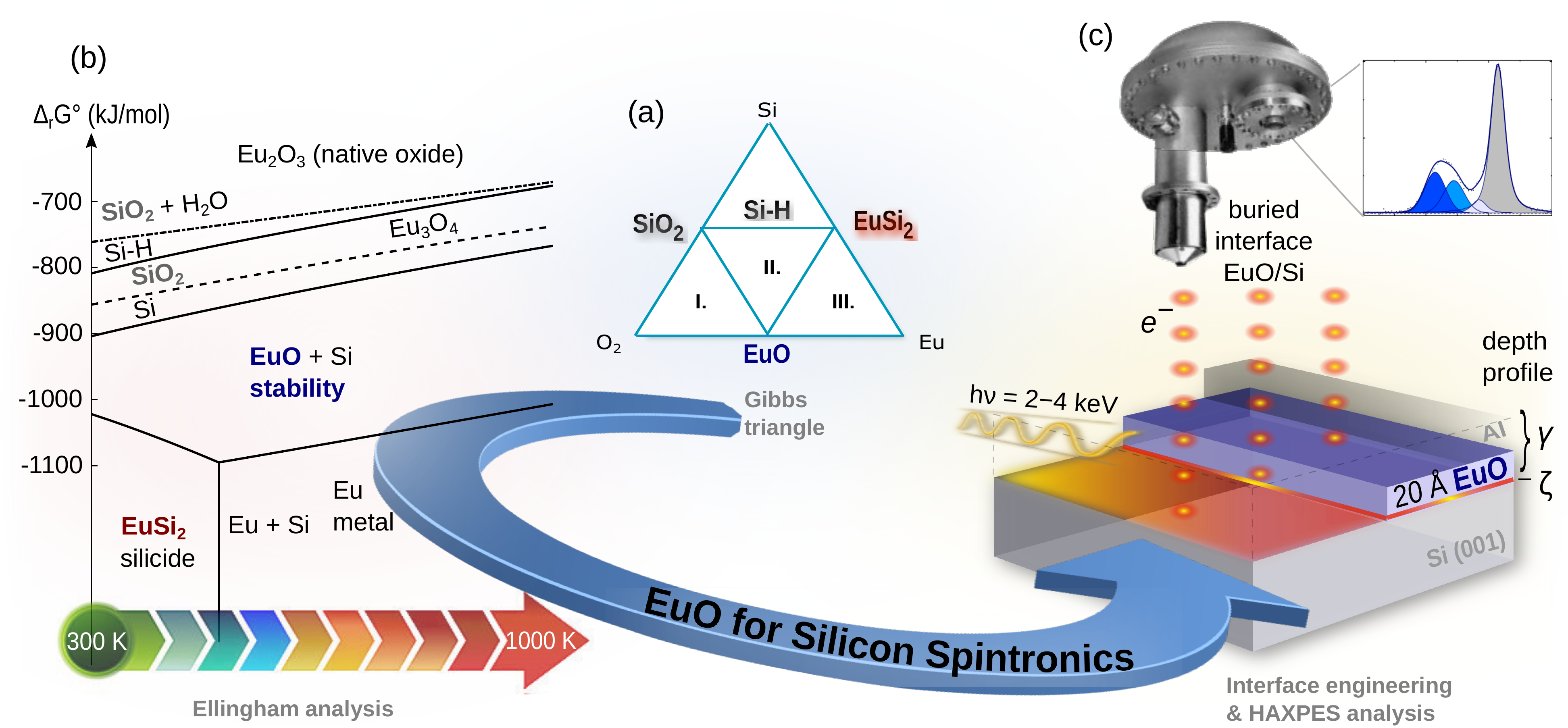}		
			\caption{Concept for the chemical and epitaxial interface engineering of ultrathin EuO directly on Si(001). (a) The three-reactant Gibbs triangle outlines the metastable EuO and reaction products of EuO synthesis on Si. (b) A thermodynamic Ellingham analysis provides EuO stability regions on Si and of the possible contaminants SiO$_2$ and EuSi$_2$. (c) Interface-passivated EuO/Si spin contacts are chemically depth-profiled using HAXPES.} 
			\label{Fig:thermodynamics-regimes}
		\end{figure*}
		
		While the synthesis and bulk properties of EuO single crystals have been investigated since the 1970s, \citet{Schmehl2007} reported on bulk-thick EuO films epitaxially grown on Si (001) with an oxide buffer in the last decade. Recently, the direct growth of polycrystalline EuO films directly on silicon (001) with tunnel barrier thickness was reported.\cite{Caspers2011a,Caspers2011b} 
		These studies proove the principle of EuO synthesis on a clean Si (001) wafer, but did not explore the spin injection interface.

		The key advantage of the spintronics system EuO/Si (001) would be the epitaxial integration with an atomically sharp interface in order to achieve a strong spin filter tunneling effect.
		Stabilizing single-crystalline EuO tunnel barriers on silicon is complicated by the formation of amorphous silicon oxides  (SiO$_{2-\delta}$)\cite{Caspers2013d} or metallic silicides (EuSi$_y$),\cite{Lettieri2002,Lettieri2003,Mundy2014,Averyanov2015} which readily form during reactive oxide synthesis at elevated substrate temperatures.[see the supplemental information] 
		Both contaminants counteract MO/Si heteroepitaxy and may reduce or even completely suppress the spin filter tunnel functionality.[see the supplemental information]

		In order to overcome these challenges, previous studies used an additional metallic\cite{Mueller2009} or oxidic buffer layer between silicon and EuO as a chemical diffusion barrier, such as SrO, BaO or MgO, which allow for a successive heteroepitaxial growth of EuO.\cite{Lettieri2003,Schmehl2007,Swartz2012,Mundy2014,Caspers2013} 
		These additional layers are several nanometers thick and have to be synthesized in a separate process step on top of the silicon substrate. 
		In spin-dependent transport, however, they act as additional tunnel barriers and have the drawback of significantly reducing the electron transfer ratio and spin filter effect. 

		Currently, the use of Eu silicides for the epitaxial integration of EuO with Silicon has gained attention. \cite{Jansen2012,Caspers2011a,Lettieri2003,Lettieri2003,Mundy2014,Averyanov2015}
		Indeed, heteroepitaxy could be shown with Eu silicides in the monolayer regime at the EuO/Si interface,\cite{Lettieri2003,Lettieri2003,Averyanov2015} however, the silicides introduce metallic conductivity and thereby suppress the spin filter functionality of the ultrathin EuO barriers.
		Thus, growing intentionally Eu silicides on Si is contrary to spin-functional silicon spintronics junctions based on the tunneling effect of magnetic oxides.
		To date, a systematic study which focusses the overall chemical cleanliness in ultrathin epitaxial EuO/Si (001) structures ---  tolerating neither additional oxides nor growing Eu silicides --- still remains elusive in order to propel the spintronic development to create a strong spin filter contact directly on silicon wafers.
			
			\begin{figure}
				\hspace*{-1mm}\includegraphics[width=0.55\columnwidth]{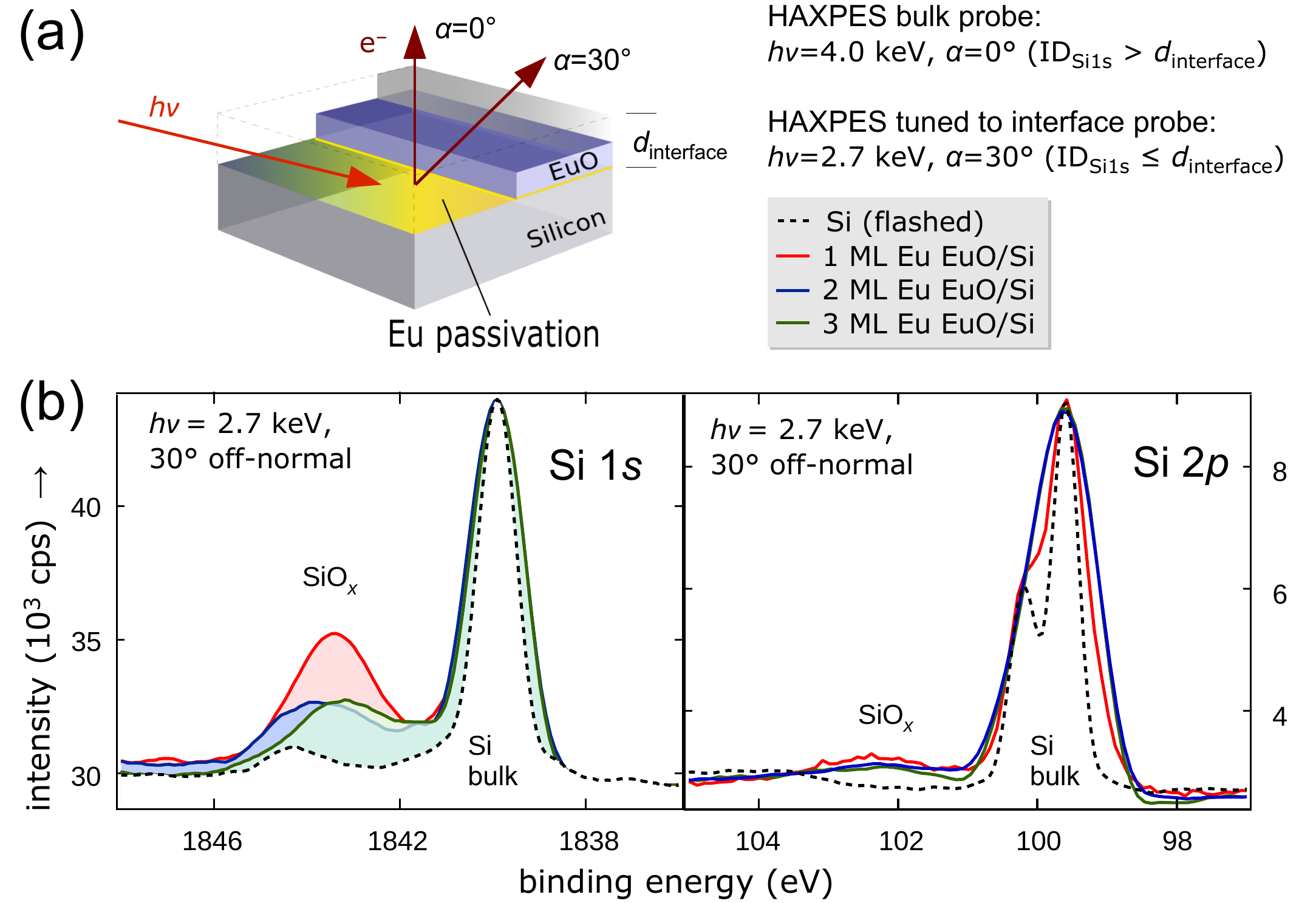}		
				\caption{Interface-sensitive HAXPES on EuO/Si (001) spin contacts for 1--3 ML Eu passivation layers on Si (001). (a) Geometrical configuration of HAXPES measurement. (b)
				Si $1s$ and Si $2p$ spectra recorded at $h\nu=\unit{2.7}{\electronvolt}$ and $\alpha=\unit{30}{\degree}$ with chemically shifted SiO$x$ contributions.} 
				\label{Fig:Si-Eu-EuO-HAXPES}
			\end{figure}
		By careful interface engineering in this study, we successfully realize the heteroepitaxial integration of EuO/Si spin contacts directly on Si (001) without any additional buffer layers. 
		The central question is how to avoid formation of silicon oxides or metallic silicides by appropriate silicon passivations. 
		While the formation and chemical stability of Si oxides is well known, we calculated the heat of formation and dissolution for the Eu silicide using density-functional pertubation theory (see supplemental).
		Our thermodynamic analysis [see the supplemental information] guided us to two complementary passivation strategies as schematically depicted in Fig.\,\ref{Fig:thermodynamics-regimes}: 
		 Passivating Si (001) with (\emph{i}) oxygen-protective Eu monolayers can avoid interfacial SiO$_x$ oxidation, and (\emph{ii}) a complete \emph{in-situ} H-passivation can prevent the formation of Eu silicides.
		 Herein, the H-Si (001) surface is expected to act like a Schottky barrier of band gap 2 eV.\cite{Yap2015}
		 In this way, spin-conserved tunneling is maintained. 
		 However, \citet{Yap2015} mention a spacial inversion of electronic surface states, hence growth and epitaxy on top of H-Si (001) need to be newly explored with respect to bare Si(001) heterostructures.
		 Both passivation methods have been mutually optimized as a function of the EuO synthesis temperature, as a fine tuning parameter for the growth kinetics at elevated temperatures.\cite{Seiler2014}
				
		In order to characterize the electronic and chemical structure of the passivated EuO/Si interfaces with tunable depth sensitivity, we use hard x-ray photoemission spectroscopy (HAXPES), which --- as the high-energy variant of soft x-ray PES --- offers  tunable and bulk-sensitive information depths $I\!D\geqslant\unit{10}{\nano\metre}$.\cite{Kobayashi2009,Fadley2010,Drube2005,Tanuma2011} 
		The obtained results unify a chemically clean interface, heteroepitaxy on Si (001) and ferromagnetism of the spin filter oxide EuO using two \emph{in situ} Si interface passivations in Oxide MBE.
		Such combined interface optimization of the magnetic oxide EuO directly on Si (001) paves the way for spin-functional EuO tunnel contacts in silicon spintronics, without the need of auxiliary buffer layers.


	\section{Results and discussion: Ultrathin \texorpdfstring{E$\boldsymbol{\mathrm{u}}$O}{EuO} on passivated \texorpdfstring{S$\boldsymbol{\mathrm{i}}$}{Si} (001)}
	\label{sec:results-HAXPES}

		In the following, we focus on the chemical and epitaxial properties of the EuO/Si (001) interface.
		We will systematically explore two \emph{in situ} Si surface passivations:
		In a first step, we address the benefits of a Eu passivation in the monolayer (ML) regime to diminish polycrystalline Si oxides (Sec.\:\ref{sec:Sioxides}).
		In a second step, we study the silicide minimization by an \emph{in situ} H-termination of the Si (001) surface (Sec.\:\ref{sec:silicides}). 
		Finally, we combine these passivation techniques in Sec.\:\ref{sec:combined} and show the impact of this careful interface engineering on the chemical structural and ferromagnetic properties of the EuO/Si spin contacts.

		\subsection{EuO/Si with Eu interface passivation}
		\label{sec:Sioxides}


			First, we aim at minimizing the formation of polycrystalline SiO$_{2-\delta}$ at the EuO/Si (001) interface, which impede heteroepitaxial growth of EuO on Si (001). 
			Among the interfacial contaminations considered in this study, SiO$_2$ is thermodynamically most stable on Si. [see the supplemental information]
			In order to render the SiO$_{2-\delta}$ formation least probable, we apply one to three ML of atomic Europium to the Si surface before EuO synthesis.


			HAXPES core-level spectra provide element-specific information of the Eu-passivated EuO/Si (001) interface. 
			We tuned the information depth to selectively probe the EuO/Si interface using photons of \unit{2.7}{\kilo\electronvolt} and off-normal electron emission  (Fig.\:\ref{Fig:Si-Eu-EuO-HAXPES}a). 
						
			Core-level photoemission spectra of Si $1s$ and $2p$ are given in Fig.\:\ref{Fig:Si-Eu-EuO-HAXPES}b. The chemical shifts towards higher binding energies unambiguously reveals the presence of interfacial SiO$_x$ contaminations. 
			In the following, the Si $1s$ deep core-levels are quantitatively analyzed, since their HAXPES probing depth matches the EuO/Si interface and the SiO$_2$ chemical shift is larger than for Si $2p$.

			\begin{figure*}
				\includegraphics[width=\linewidth]{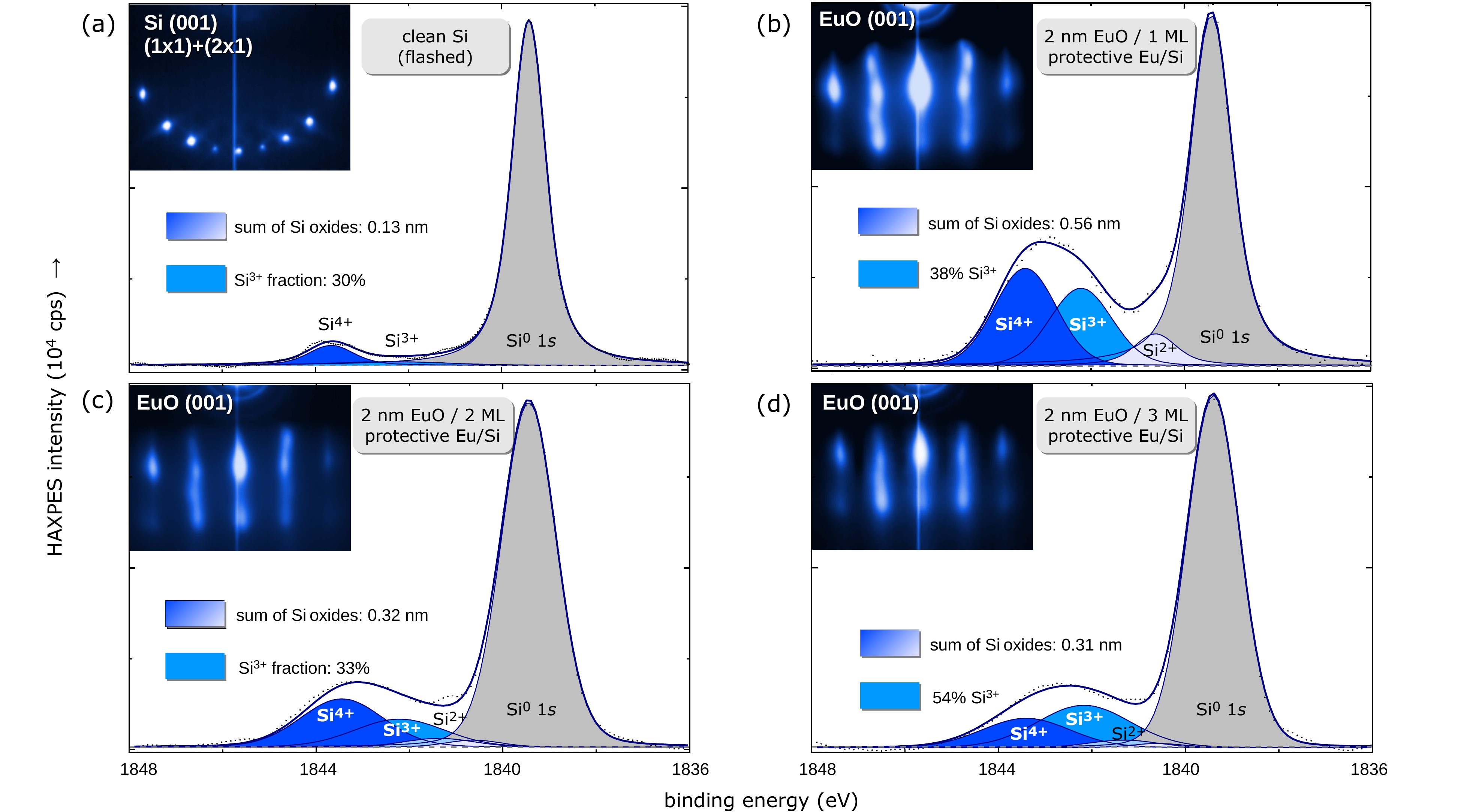}		
				\caption{Quantitative analysis of SiO$_{2-\delta}$ contaminations for different atomic Eu (1--3 ML) passivations at the 2 nm-EuO/Si (001) interface. (Insets) Corresponding RHEED patterns along [010] of the EuO film.} 
				\label{Fig:Sioxides-RHEED}
			\end{figure*}	
			\begin{figure}[t]
				\centering
				\includegraphics[width=0.51\linewidth]{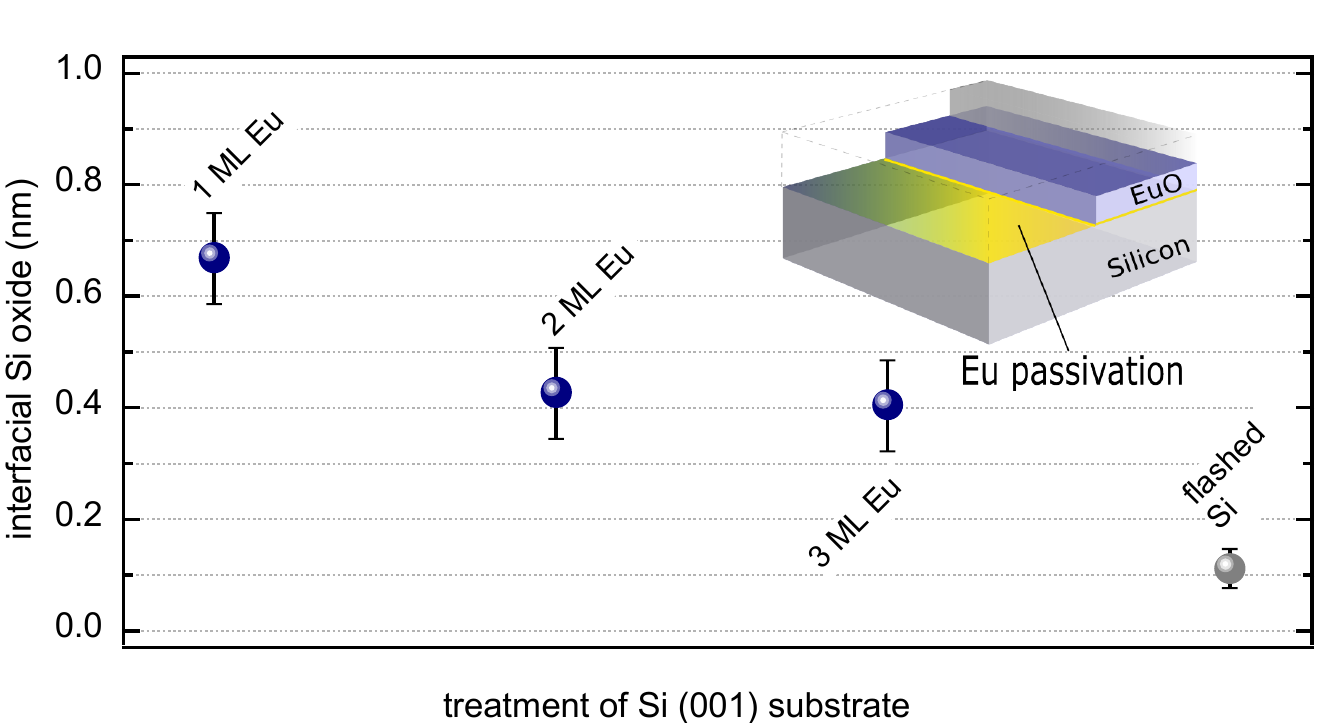}		
				\caption{Thickness of the interfacial SiO$_{2-\delta}$ contaminations as detemined from a quantitative peak fitting analysis of the Si $1s$ HAXPES spectra.} 
				\label{Fig:Sioxide-quantitative}
			\end{figure}
			The evaluation of the chemical shifts of the interfacial SiO$_{2-\delta}$ reveals the presence of different Si oxidation
states at the EuO/Si interface ($\delta=0..1$). 
			The fraction of Si$^{3+}$ valency ranges from 33\% to 38\% for one and two ML Eu passivation, respectively.
			A maximal Si$^{3+}$ fraction of 54\% is the result of three ML protective Eu.
			The fractional under-oxidation of the silicon oxide to the valency Si$^{3+}$\! --- rather than Si$^{4+}$ for pristine SiO$_2$ --- therefore is directly related to the thickness of the Eu passivation layers.

			In order to quantify the interfacial silicon oxide formation as a function of the applied Eu interface passivation, we apply a least-squares peak fitting analysis after Levenberg-Marquardt for the three EuO/Si spin contacts.
			One ML Eu passivation reveals \unit{0.67\pm0.09}{\nano\metre} SiO$_{2-\delta}$, and a minimum of  \unit{0.42\pm0.08}{\nano\metre} SiO$_{2-\delta}$ for 3 ML Eu passivation was achieved (Figs.\:\ref{Fig:Sioxides-RHEED}, \ref{Fig:Sioxide-quantitative}). 
			A tiny fraction of residual SiO$_2$ (\unit{0.11\pm0.03}{\nano\metre}) from the flashed silicon wafers appeared to be systematically present in
all samples.

			\begin{figure}
				\includegraphics[width=0.51\linewidth]{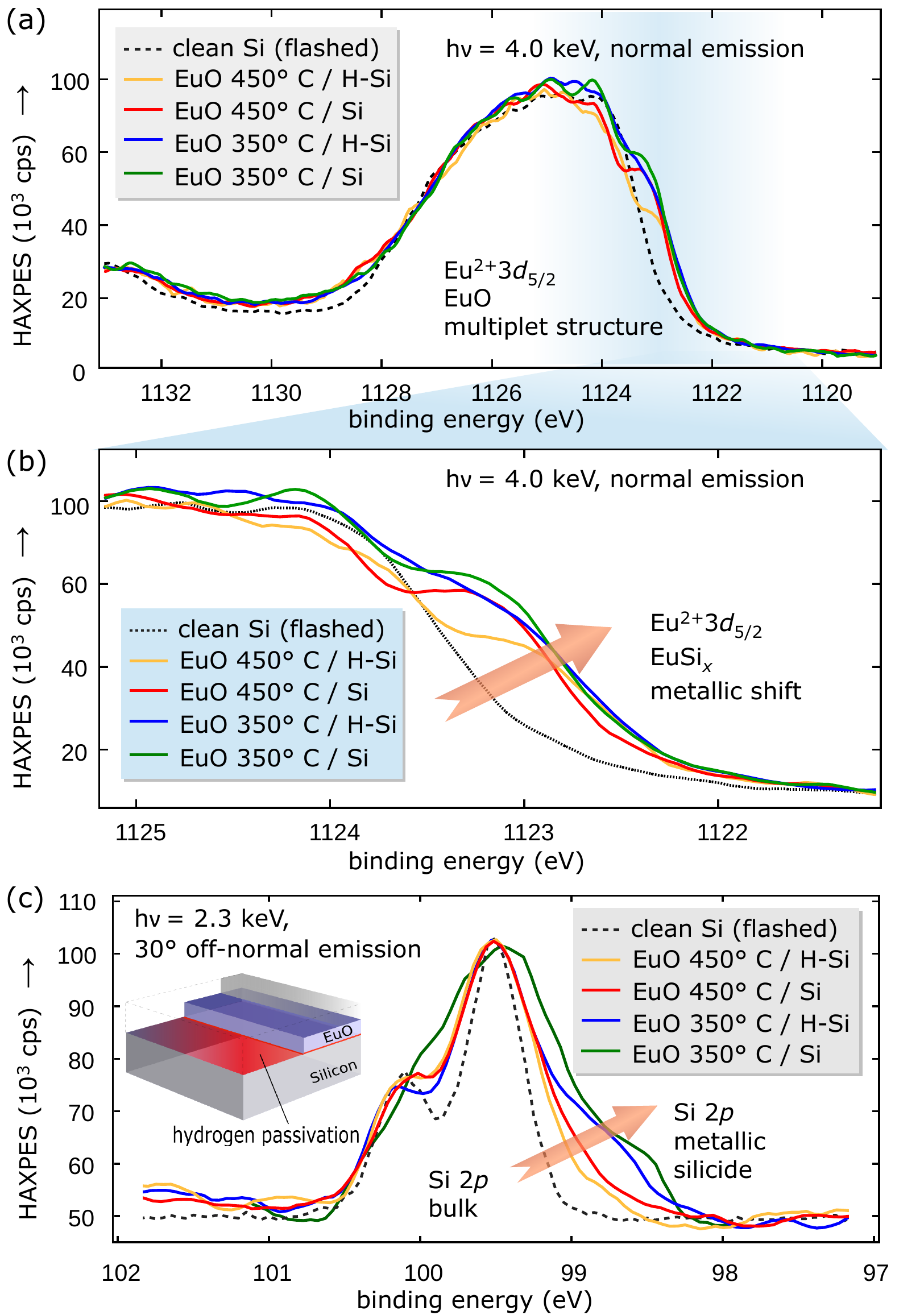}
				\caption{(a), (b) HAXPES of the Eu $3d_{\nicefrac{5}{2}}$ multiplet structure of H-terminated EuO/Si (001) spin contacts grown at different substrate temperatures. (c) Si $2p$ spectra  with the photon energy tuned to selectively probe the EuO/Si interface.} 
				\label{Fig:silicides-qualitative}
			\end{figure}				
			In the following, we focus on the structural properties of the Eu-passivated EuO/Si (001) heterostructures.
			The crystallinity of ultrathin EuO (2 nm) is monitored by electron diffraction (RHEED in Fig.\:\ref{Fig:Sioxides-RHEED}) along the [010] direction. 
			The flashed Si (001) reference surface exhibits a clear $(1\times 1)$ bulk diamond structure and equally intense Si (001) $(2\times 1)$ pattern of the surface reconstruction (Fig.\:\ref{Fig:Sioxides-RHEED}a).\\ 
			The 2 nm EuO/Si (001) sample with 1 ML Eu passivation reveals linear patterns of the EuO \emph{fcc} lattice (Fig.\:\ref{Fig:Sioxides-RHEED}b), whereby the rods have a different distance for alternating vertical diffraction orders. 
			The analysis of the diffraction rods shows a stepwise relaxation of the EuO lattice to its native parameter towards more grazing diffraction.
			The RHEED pattern for EuO/Si with 2 ML and 3 ML oxygen-protective Eu passivation also shows linear rods of crystalline EuO (Fig.\:\ref{Fig:Sioxides-RHEED}b). 
			Here, EuO adapts the Si (001) lattice parameter heteroepitaxially and shows only a negligible lateral relaxation within the 2 nm EuO film thickness.
			We conclude, that epitaxial growth of EuO on Eu-passivated Si (001) is feasible, where EuO adapts its native lattice parameter gradually in samples with 1 or 2 ML protective Eu at the EuO/Si (001) interface. 
			A heteroepitaxial relation of EuO to Si (001) without observable strain relaxation is possible with 2 ML of atomic Eu passivation.




		\subsection{EuO/Si with hydrogen interface passivation}
		\label{sec:silicides}
		
			As a second step, we investigate the formation of interfacial silicides, which both causes polycrystalline EuO growth and introduces metallic conductivity at the interface --- with detrimental effect on any tunnel transport. 
			Consequently, in order to sustain electrically insulating properties, silicide formation must be minimized such that no conductive spots are introduced in the EuO/Si spin contacts.
			Based on the Ellingham analysis (Fig.\:\ref{Fig:thermodynamics-regimes}),[see the supplemental information] we apply an \emph{in situ} hydrogen-passivation to the Si (001) surface in order to create a EuO/Si interface with minimal silicides. 
			As a second control parameter, the substrate temperature $T_{\mathrm{S}}$, which directly controls the EuO distillation kinetics,\cite{Altendorf2011} is varied.

			\begin{figure*}
				\includegraphics[width=\linewidth]{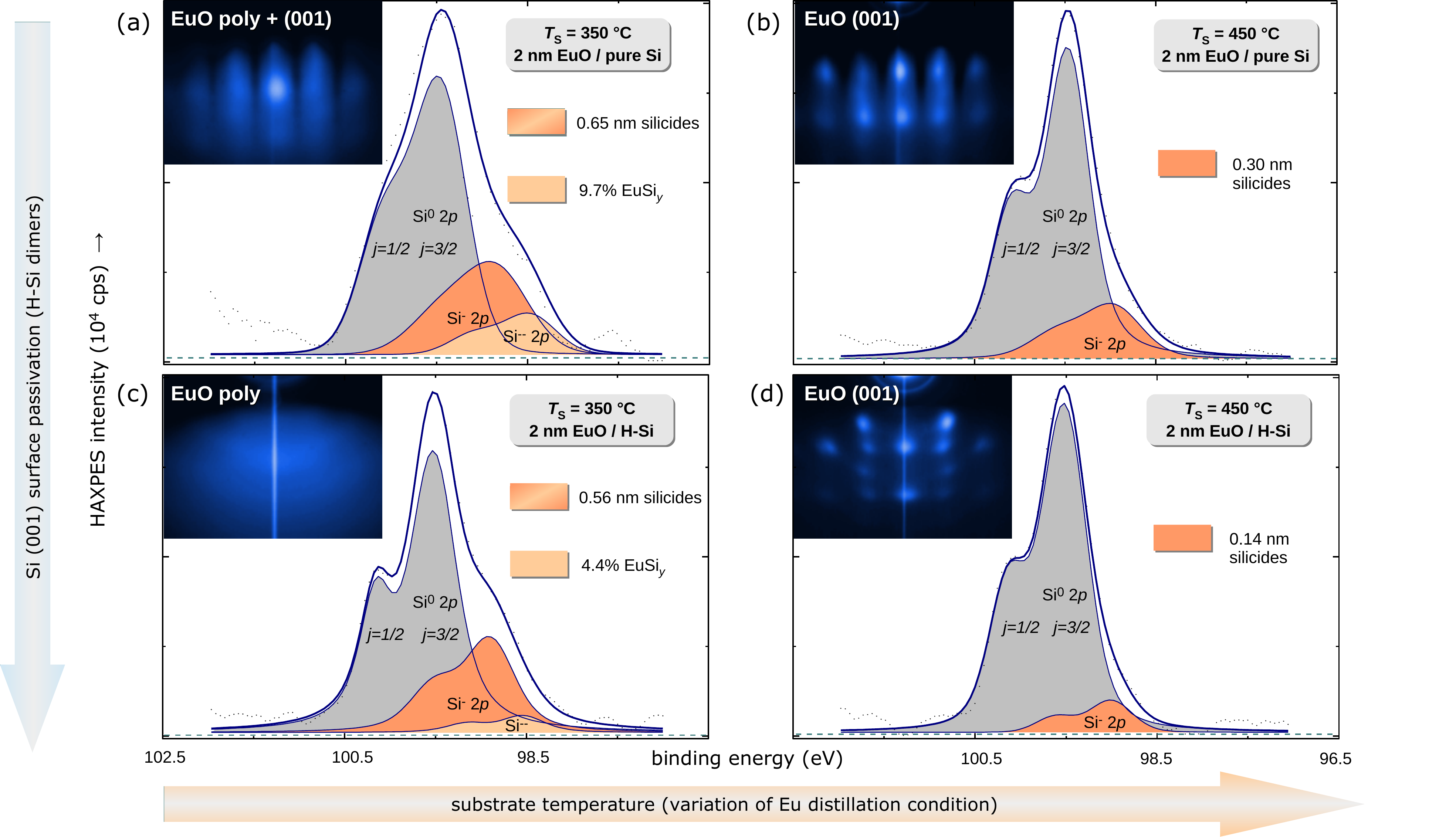}	
				\caption{Quantitative analysis of the Si $2p$ peak. The model involves two silicide phases which are highlighted. RHEED patterns of the EuO/Si spin contacts after 2 nm EuO synthesis has finished are inset. The matrix of parameters diplayed contains a variation of the substrate temperature (horizontal) and the hydrogen termination of silicon (vertical).} 
				\label{Fig:silicides-fits}
			\end{figure*}
			\begin{figure}[b]
				\centering
				\includegraphics[width=0.55\linewidth]{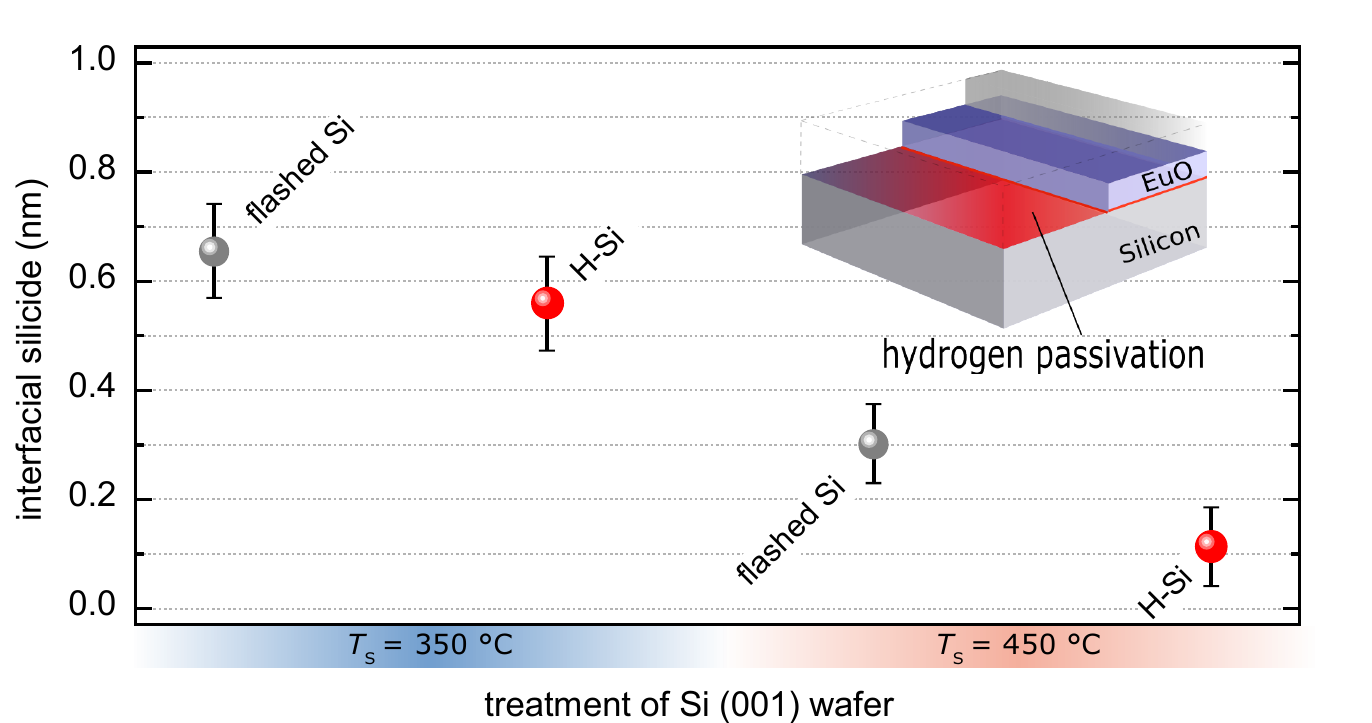}		
				\caption{Thickness of the interfacial silicide contamination determined from a Si $2p$ peak fit analysis for H-passivated Si (001) and varied substrate temperature.} 
				\label{Fig:silicides-quantitative}
			\end{figure}		

			First, we study the chemical properties of the buried EuO/H-Si heterointerface by analysis of the Eu and Si core levels.
			The Eu $3d_{\nicefrac{5}{2}}$ HAXPES spectra (Fig.\:\ref{Fig:silicides-qualitative}a) provide information of the ultrathin EuO layer adjacent to the Si interface.
			The Eu silicide component is located at the low binding energy side of the spectra, respectively, and its spectral weight systematically varies with the different H-passivation conditions. 
			Because the Eu $3d$ core level (Fig.\,\ref{Fig:silicides-qualitative}a,b) exhibits a rather complex multiplet structure,\cite{Caspers2011a} we chose the Si $2p$ core level (Fig.\,\ref{Fig:silicides-qualitative}c) for quantitative analysis. 
			As shown in Fig.\:\ref{Fig:silicides-qualitative}c, the metallic silicide component is chemically shifted by \unit{-0.6}{\electronvolt} to the low-binding energy side of the bulk Si $2p$ doublet. 
			A clear dependence is observable: Both an increase of the substrate temperature $T_S$ as well as applying the H-passivation of Si (001) reduces the silicide fraction at the EuO/Si interface.
			Thereby, the effect of increasing the EuO synthesis temperature from \unit{350}{\celsius} to \unit{450}{\celsius} is larger ($-38\%$ silicide) than the application of hydrogen passivation ($-14\%$ silicide) of the Si (001) surface (Figs.\:\ref{Fig:silicides-fits}, \ref{Fig:silicides-quantitative}). 
			The benefit of a higher EuO synthesis temperature origins from the thermodynamic reaction balances of the EuSi$_2$ Gibbs free energy of formation:[see the supplemental information] the europium silicide favorably dissolves with nearly temperature-independent probability during EuO distillation growth, this yields EuO and Eu.
			At the same time, any excess Eu is increasingly re-evaporated at higher $T_{\mathrm{S}}$ due to Eu distillation, and this efficiently minimizes the available constituents for silicide formation.  
			The advantage of a hydrogen passivation, although smaller in effect compared to the temperature variation, also origins from its higher thermodynamic interface stability (as evidenced by HR-TEM in \citet{elsewhere2014}) compared to the silicide (Fig.\:\ref{Fig:thermodynamics-regimes}). \\
			We conclude at this point, that suitable parameters for a minimized interfacial silicide formation ($d(\mathrm{EuSi_2})\approx\unit{0.14}{\nano\meter}$) are the application of high EuO synthesis temperatures ($\geqslant\unit{450}{\celsius}$) and a complete hydrogen passivation of the Si surface.
			
			\begin{figure*}
				\centering\includegraphics[width=0.98\linewidth]{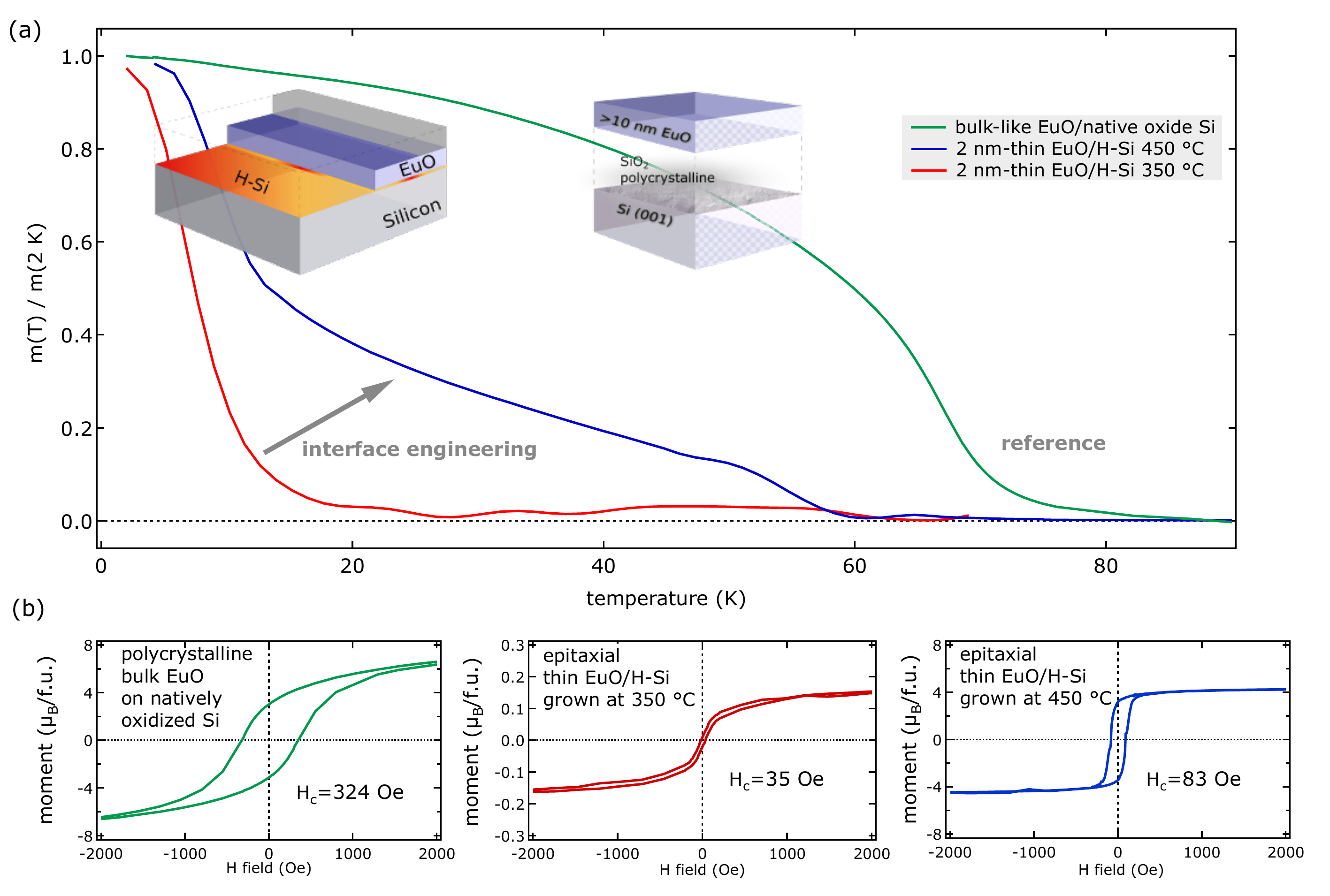}	
				\caption{Ferromagnetic properties of ultrathin EuO on H-passivated Si (001) and of a bulk EuO reference grown on native SiO$_2$/Si. (a) Magnetization vs. $T$ recorded at saturation field. Panel (b) shows hysteresis loops $M(H)$ recorded at $T=\unit{2}{\kelvin}$. The formula unit (f.u.) used for normalization is one Eu$^{2+}$O$^{2-}$\,in an \emph{fcc} lattice with cubic parameter $a=\unit{5.14}{\angstrom}$.} 
				\label{Fig:silicides-magnetic}
			\end{figure*}			

			Next, we investigate the crystalline structure of the EuO/Si heterostructures dependent on the H passivation of Si (001) and the EuO synthesis temperature $T_{\mathrm{S}}$. 
			Figure\:\ref{Fig:silicides-fits}a-d depict the RHEED pattern after growth of the \unit{2}{\nano\metre} ultrathin EuO.
			At lower synthesis temperature ($T_{\mathrm{S}}=\unit{350}{\celsius}$), we observe weak RHEED pattern of the cubic EuO \emph{fcc} structure only if EuO is directly deposited on Si (001) (inset in Fig.\:\ref{Fig:silicides-fits}a).
			This low crystallinity is explained by the presence of excess Eu and silicides from an incomplete Eu distillation condition (Fig.\:\ref{Fig:silicides-fits}a, c).
			At elevayed substrate temperature ($T_{\mathrm{S}}=\unit{450}{\celsius}$), in contrast, single-crystalline EuO with sharp-defined reciprocal rods in the RHEED pattern are observed. 
			In this synthesis regime, the application of a H-passivation to the Si surface facilitates an highly ordered EuO growth, but, at the same time, introduces a certain roughness in growth direction as reflected by the distinct vertical maxima in the RHEED pattern in Fig.\:\ref{Fig:silicides-fits}d, which we attribute to different H-Si surface reconstructions.\cite{Northrup1991,Hansen1998}
			Thus, the result is twofold: 
			We find the smoother EuO film at high synthesis temperature ($T_{\mathrm{S}}=\unit{450}{\celsius}$) which confirms heteroepitaxy of EuO directly on flashed Si (001). 
			A sustained crystalline relationship with sharper diffraction pattern of EuO, however, is formed at higher EuO synthesis temperature ($T_{\mathrm{S}}=\unit{450}{\celsius}$) and a complete H-passivation of Si (001).

			In the following, we explore the ferromagnetic properties of ultrathin magnetic oxide EuO on H-passivated Si (001), as important prerequisites for creating a strong spin filter contact.
			We compare ultrathin heteroepitaxial EuO on H-passivated Si (001) at lower and higher synthesis temperatures with a bulk-like EuO/Si reference sample in Fig.\:\ref{Fig:silicides-magnetic}.
			The ultrathin epitaxial EuO spin contacts are ferromagnetic.
			However, the Curie temperature $T_C$ and the magnetic saturation moment $M_S$ both show a strong dependence on the interface passivation.
			For 2 nm EuO/H-Si synthesized at lower temperature, $T_{\mathrm{C}}$ is dramatically reduced to $\approx\unit{10}{\kelvin}$ (Fig.\:\ref{Fig:silicides-magnetic}a) and the field dependence shows mainly paramagnetic behavior (Fig.\:\ref{Fig:silicides-magnetic}c).
			Consistent with the HAXPES analysis, a coverage of paramagnetic europium silicides at the EuO/Si interface results in this magnetic response.
			In contrast,  the Curie temperature of the high-$T_{\mathrm{S}}$ EuO spin contact is bulk-near $\approx\unit{58}{\kelvin}$, which agrees well with results obtained for the reference sample.
			Hence, the reduction of the Curie temperature is explained by a reduced nearest neighbor (NN) interaction in laterally expanded ultrathin EuO films\cite{Mueller2009b,Caspers2013} rather than to formation of silicides at the interface. 
			Also, the field dependence of the high-$T_{\mathrm{S}}$ EuO spin contact (Fig.\:\ref{Fig:silicides-magnetic}b) agrees well with a successful diminishment of the paramagnetic silicides, with an emergent ferromagnetic switching behavior and a bulk-near specific magnetic moment ($M_{\mathrm{S}}=4.4\,\mu_{\mathrm{B}}/\mathrm{Eu^{2+}}$; $M_{\mathrm{S}}^{\mathrm{bulk}}=7\,\mu_{\mathrm{B}}/\mathrm{Eu^{2+}}$).

		\subsection{Combined interface passivation}
		\label{sec:combined}
			In a last step, we ultimately combine the benefits of the two complementary  EuO/Si interface passivations:
			(\emph{i}) the hydrogen-passivated Si (001) surface and (\emph{ii})  the Eu monolayer-passivated Si interface.
			By integrating these two \emph{in situ} passivation techniques we simultaneously minimize Eu silicides as well as SiO$_{2-\delta}$ at the EuO/Si (001) interface --- but the challenge remains to find the optimal thermodynamic conditions.

			As a starting point, we use the optimum set of passivation parameters of the previous sections for the silicide as well as the silicon oxide minimization: 
			the EuO/Si (001) heterointerface is now passivated with two monolayers of Eu in addition to a complete \emph{in situ} H passivation of the Si (100) wafer. 
			The temperature of EuO synthesis $T_{\mathrm{S}}$  is the remaining variable, controlling the chemical reactivity and growth kinetics during EuO synthesis. 
			We vary $T_{\mathrm{S}}$ within the Eu distillation condition\cite{Altendorf2011} around $T_{\mathrm{S}}=\unit{450}{\celsius}$, at which an optimized crystalline quality of EuO was observed before, and investigate three interface-passivated EuO/Si (001) spin contacts prepared at $T_{\mathrm{S}}=\unit{400}{\celsius},\ \unit{450}{\celsius}\text{ and }\unit{500}{\celsius}$.

\begin{figure}
				\centering
				\hspace*{-2mm}\includegraphics[width=0.55\linewidth]{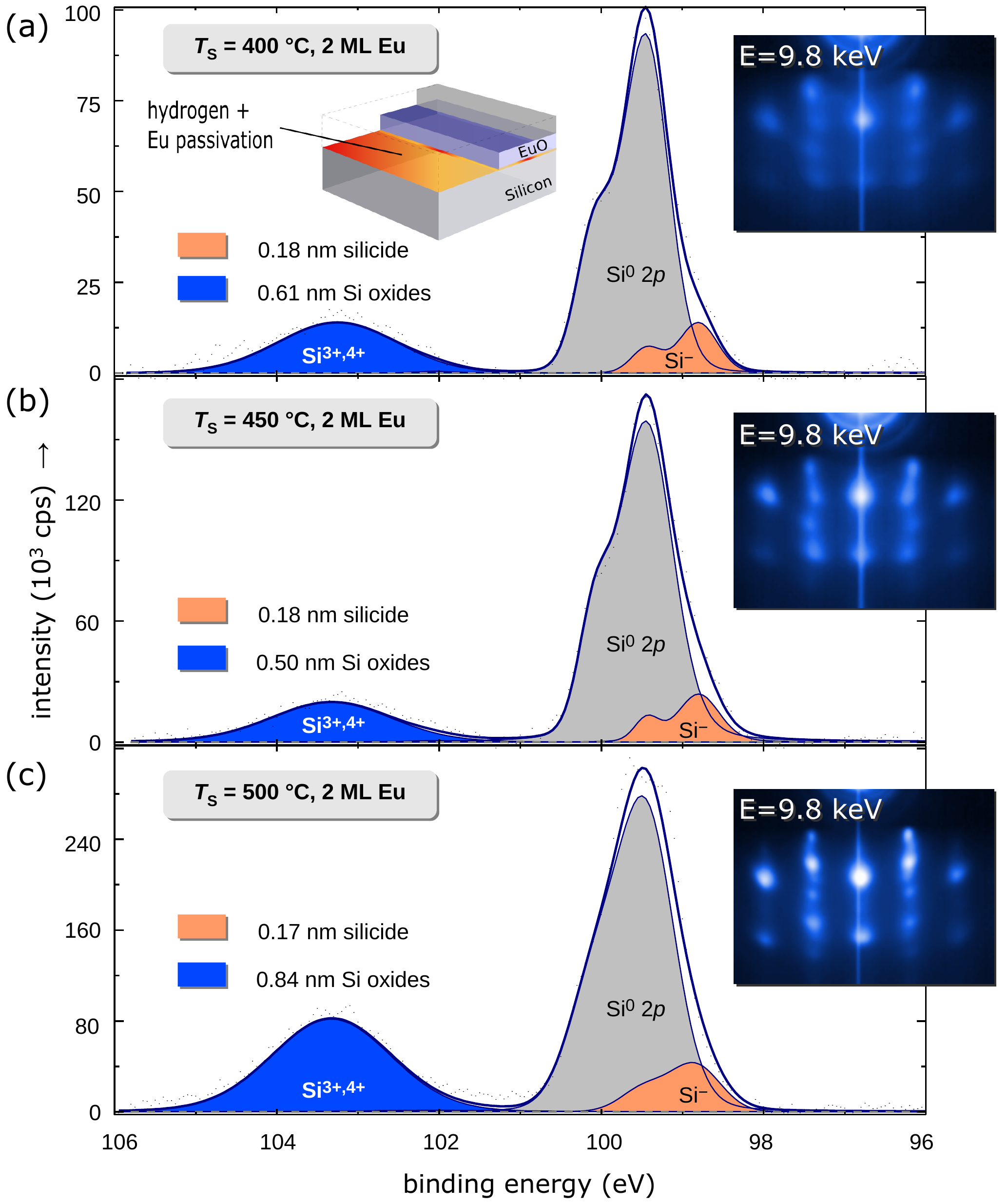}		
				\caption{HAXPES Si $2p$ spectra for quantitative chemical analysis of the combined Eu- and H-Si passivation of the EuO/Si (001) interface for varying $T_S$ EuO distillation growth. The insets show the respective RHEED pattern for 2 nm EuO/Si (001).}
				\label{Fig:combined}
			\end{figure}

			Consistent with the previous optimizations, we quantify the chemical state of the EuO/Si heterointerface with combined H-Si and Eu monolayer passivation by interface-sensitive HAXPES analyses of the Si $2p$ spectra (Fig.\:\ref{Fig:combined}).
			As a first result we note that varying $T_S$ for EuO synthesis does not show a significant effect on the formation of Eu silicides: 
			the EuSi$_2$ interfacial thickness is about $\approx$\unit{0.2}{\nano\metre} for all three samples, in good agreement with the optimum value of the silicide optimization study (Sec.\,\emph{EuO/Si with hydrogen interface passivation}). 
			The EuO/Si interface oxidation, however, exhibits a clear minimum value of $\approx$\unit{0.69}{\nano\metre} for the sample synthesized at $T_{\mathrm{S}}=\unit{450}{\celsius}$, while a maximum oxidation of the Si interface is observed for the EuO/Si interface which was exposed to the highest EuO synthesis temperature, $T_{\mathrm{S}}=\unit{500}{\celsius}$. 
			We attribute the enhanced Si oxidation to an increased Si mobility and reevaporation of the Eu interface passivation at the highest $T_{\mathrm{S}}$.

			We correlate the chemical interface properties directly with the crystalline structure of the ultrathin EuO layers (see insets in Fig.\:\ref{Fig:combined}). 
			The RHEED pattern indicate a well-defined EuO single-crystalline structure at both higher synthesis temperatures, $T_{\mathrm{S}}=\unit{\text{450--500}}{\celsius}$. 
			We ascribe this improved heteroepitaxy to a complete Eu distillation process and the consequent suppression of EuSi$_y$ formation, in accordance with the silicide optimization study (Sec.\,\emph{EuO/Si with hydrogen interface passivation}). 
			However, at highest synthesis temperature $T_{\mathrm{S}}=\unit{500}{\celsius}$ a significant three-dimensional growth was indicated by electron diffraction, likely due to faster growth kinetics and enhanced Si oxidation. 
			All three EuO samples in this combind Si oxide + Eu silicde optimization study confirm a clear cubic heteroepitaxy on passivated Si (001) without signs of strain relaxation.
			
			The ferromagnetism of these optimal EuO spin contacts was confirmed up to \unit{58}{\kelvin} by SQUID magnetometry and all resemble the same $M$ vs $T$ behavior as shown in Fig.\:\ref{Fig:silicides-magnetic}a, which agrees well with the silicide diminishment to the same minimum.

		\subsection{Summary of the EuO/Si (001) interface optimization studies}
		\label{sec:summary}
\begin{figure*}
				\centering
				\includegraphics[width=\linewidth]{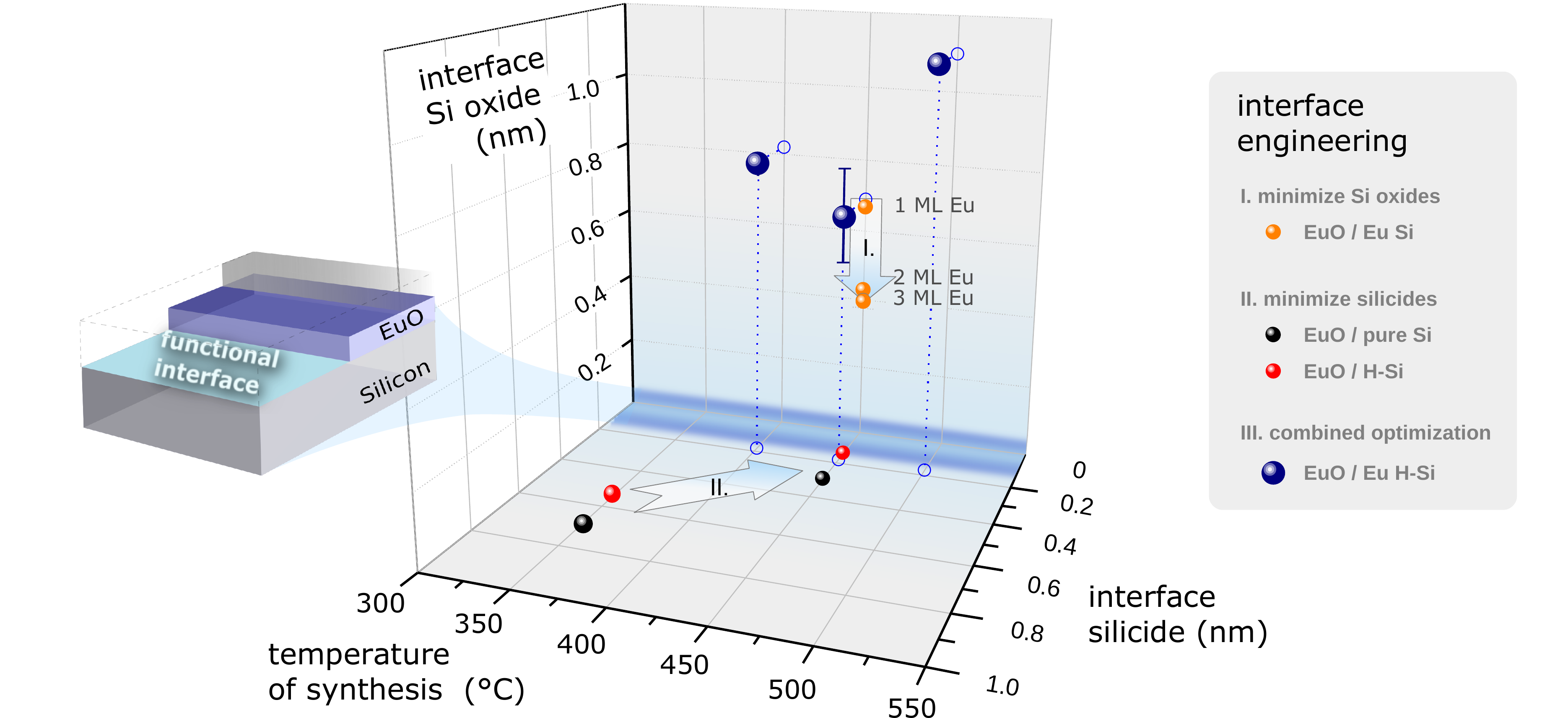}		
				\caption{Direct correlation between the successive Eu/Si (001) interface optimization steps (i) Eu-passivation, (ii) H-Passivation and (iii) combined Eu- and H-passivation at different EuO distillation temperatures $T_S$., for which the Eu silicides EuSi$_2$ and  SiO$_{2-\delta}$ interface contamination is systematically minimized.} 
				\label{Fig:summary3d}
			\end{figure*}

	                  Three complementary optimization studies of the EuO/Si (001) interface correlating chemical (HAXPES), structural (LEED, RHEED) and magnetic (SQUID) characteristics were conducted based on atomic Eu- and H-Si (001) passivations, respectively. 
	                  As a function of these interface passivations, we evaluated the Si $1s$, $2p$ and Eu $3d$ core level spectra for quantification of the Si oxides SiO$_{2-\delta}$ and Eu silicides EuSi$_2$ at the buried EuO/Si interfaces.
			In Fig.\:\ref{Fig:summary3d}, we summarize the results of all successive interface passivation steps, for which EuO formed on Si (001) in a clear heteroepitaxial relation: 
			The optimal parameter set (blue spheres in Fig.\:\ref{Fig:summary3d}) was achieved using a combined Si surface passivation of (\emph{i}) complete H-Si and (\emph{ii}) 2 ML of protective Eu. 
			A minimum of both oxidic and metallic contaminations was reached at the synthesis temperature with optimal Eu distillation growth ($T_{\mathrm{S}}=\unit{450}{\celsius}$), which also revealed the highest quality single-crystallinity of the ultrathin EuO.			
		
	\section{Conclusions}
	\label{sec:conclusion}
	
		EuO on Silicon (001) is perfectly suited to create strong spin filter contacts for silicon spintronics, but the oxide-semiconductor interface notoriously suffers from chemical reactions.
		We succeeded in integrating ultrathin heteroepitaxial EuO on Silicon without additional oxide buffer layers. 
		Our thermodynamic analysis guided us to select two complementary Si passivation techniques: an in-situ hydrogen passivation of the Si (001) surface to avoid metallic silicides, and oxygen-protective Eu monolayers in order to effectively suppress SiO$_x$ species at the interface.

		We performed a systematic study on the beneficial impact of each passivation step and finally combined the results to gain the optimum set of synthesis conditions for EuO/Si (001) spin contacts.		
		The interface contamination was reduced to a sub-monolayer coverage of $\sim$\unit{0.14}{\nano\metre} Eu silicides at the H-passivated EuO/Si interface and a minimum of $\sim$\unit{0.42}{\nano\metre} interfacial Si oxides. 
		
		With combined atomic Eu- and H-Si (001) interface passivations, we succeeded in synthesizing well-defined  2\,nm EuO single-crystalline films with  ferromagnetic properties  up to $T_{\mathrm{C}}=\unit{60}{\kelvin}$.

		The effective \emph{in situ} passivation techniques proposed in this study are a key to quench Eu silicide and Si oxide species at EuO/Si heterointerfaces down to the submonolayer regime at the same time, which in turn enables  EuO/Si (001) heteroepitaxy with sizeable ferromagnetic properties.
		Alternative routes to achieve silicide-free EuO interfaces on silicon are pointed out in the supplemental information.[see the supplemental information]
		Such complementing passivation methods permit the successful interface engineering of the magnetic insulator EuO directly with Silicon wafers --- paving the way for a seamless integration of spin-functional magnetic oxides with Silicon nanotechnology.
\bigskip

	\section{Experimental Methods}
		\label{sec:experimental}
	
		Si (001) wafer pieces were cleaned by  \emph{in situ} thermal flashing at 1200$^{\circ}$C, which resulted in the typical ($2\times 1$) surface reconstruction as observed by low energy electron diffraction (LEED).  
		\emph{In situ} and right before EuO synthesis, surface passivation which are subject to this study were applied.
		Either monolayers of sublimed Eu or atomic hydrogen through a tungsten gas cracker were supplied to the sample surface.	
		
		In order to synthesize stoichiometric EuO ultrathin films ($d_{\mathrm{EuO}}=\unit{20}{\angstrom}$), we applied the Eu distillation condition\cite{Steeneken2002,Sutarto2009,Altendorf2011} in ultrahigh vacuum with base pressure better than \unit{1 \times 10^{-10}}{\milli\bbar} using reactive oxygen molecular beam epitaxy (Oxide MBE) as described in \citet{Caspers2013d}.
		The EuO synthesis parameters were defined by a constant flux of Eu metal atoms  ($\Phi_{\mathrm{Eu}}=\unit{2.8\times 10^{13}}{\mathrm{atoms}\per\centi\metre\squared\usk\second}$) provided by e-beam evaporation of an Eu ingot (99.99\%), a constant supply of molecular oxygen (99.999\%) with a partial pressure of \unit{1.5 \times 10^{-9}}{\mathrm{Torr}},\cite{Caspers2013} and a varying substrate temperature $T_{\mathrm{S}}$ within the range of Eu distillation growth (\unit{\text{350--450}}{\celsius}), which in combination define the thermodynamic conditions of the EuO/Si (001) interface. 
		Eu-rich distillation growth mode is known to wet the interface very well and to yield smooth EuO films. 
		However, the change in electronic structure in H-Si or oxygen-rich EuO growth conditions could possibly yield island growth modes. 
		Our study focuses thus on the well-known Eu distillation growth mode.
		During synthesis, the evolving surface crystalline structure was monitored by reflection high-energy electron diffraction (RHEED). Finally, the EuO/Si(001) heterostructures were capped with \unit{5}{\nano\metre} protective Al.

		The chemical properties of the EuO/Si(001) interface were studied at the undulator beamline P09  {PETRA\,III (DESY, Hamburg)}\cite{Gloskovskii2012} and at the KMC-1 beamline (BESSY II, Berlin) with the high-energy endstation HIKE\cite{Gorgoi2009,Schaefers2007}  
		In order to optimize the interface sensitivity, core level spectra were acquired by angle-de\-pen\-dent HAXPES allowing us to tune the photoelectron escape depth by rotating the sample around an axis vertical to the plane of x-ray incidence and photoelectron acceptance. 
		The total energy resolution was mainly determined by the bandpass of the Si (111) double crystal monochromator corresponding to \unit{\text{300}}{\milli\electronvolt} for the excitation energy range \unit{\text{2.3--4}}{\kilo\electronvolt}, 
and all spectra were recorded at room temperature.
		Backgrounds described by the Tougaard equation\cite{Tougaard1988} were subtracted from the raw spectra to account for inelastic scattering of the photoelectrons. 		
				
		In order to quantify the thickness $\zeta$ of a possible reaction layer (rl) at the EuO/Si interface, we evaluate the spectral weights $I$ of the different chemical components in the Si spectra.
		Dependent on the depth of the buried Si interface and the overlayer thickness $\gamma$, we model the exponentially damped intensities as
		\begin{align}
				I_{\mathrm{rl}}(\zeta,\gamma,\lambda)&=e^{-\frac{\gamma}{\lambda_{\mathrm{cap}}\cos\alpha}}\;\int_\gamma^{\gamma+\zeta}e^{-\frac{x}{\lambda_{\mathrm{rl}}\cos\alpha}}\,\mathrm{d}x,\\
				I_{\mathrm{Si}}(\zeta,\gamma,\lambda)&=e^{-\frac{\zeta+\gamma}{\lambda_{\mathrm{cap}}\cos\alpha}}\;\int_{\gamma+\zeta}^\infty e^{-\frac{x}{\lambda_{\mathrm{Si}}\cos\alpha}}\,\mathrm{d}x,
							\label{eq:Irl}
			\end{align} 

		where $\alpha$ denotes the off-normal exit angle and $\lambda$ the inelastic mean free path of the photoelectrons.\cite{Tanuma2011}
		The mean depth of photoelectron emission varies as $\cos(\alpha)$,\cite{Fadley2008} and is proportional to the kinetic energy of the photoelectrons.
		The fraction $f_{\mathrm{rl}}$ of spectral contribution from the interfacial reaction layer with respect to the bulk Si spectral intensity denotes as
		\begin{align}
			\frac{I_{\mathrm{rl}}(\zeta,\gamma,\lambda)}{I_{\mathrm{Si}}(\zeta,\gamma,\lambda)}\stackrel{\text{calc}}{=}f_{\mathrm{rl}}&\stackrel{\text{meas}}{=}\frac{I_{\mathrm{rl}}}{I_{\mathrm{Si}}}\cdot\frac{\lambda_{\mathrm{Si}}n_{\mathrm{Si}}}{\lambda_{\mathrm{rl}}n_{\mathrm{rl}}}.
			\label{eq:frl}
		\end{align}
		Using $\lambda_{\mathrm{rl}}\approx\lambda_{\mathrm{Si}}$ leads to
		\begin{align}
			\zeta&=\lambda_{\mathrm{Si}}\cos\alpha\cdot\ln(f_{\mathrm{rl}}+1),
			\label{eq:c}
		\end{align}		
		with the thickness $\zeta$ of the reaction layer, which allows us to quantify the EuSi$_2$ and SiO$_{2-\delta}$ formation at the buried EuO/Si(001) interface. 
		In practice, we tune the information depth $I\!D\approx 3\lambda$ to match $\gamma+\zeta$ in order to achieve an optimal interface sensitivity.
		Calibrations of $\lambda$ are obtained from the layer thicknesses of EuO and the capping.
		Error bars in the results of $\zeta$ originate from $\chi^2$ of least-squares fitting analyses of the HAXPES spectra.

\pagebreak
\footnotesize
		We acknowledge experimental support by S.\,D\"oring and S.\,D.\,Flade.
		Funding by the Federal Ministry of Education and Research (BMBF) under contracts 05KS7UM1,  05K10UMA, 05KS7WW3, and 05K10WW1 is gratefully acknowledged.
		ML and KZR acknowledge the funding of the Helmholtz Gemeinschaft, grant VH-NG-409 and the support of the Jülich Supercomputing Centre, grant Jiff38.		
		MM acknowledges financial support by DFG under grant MU\-3160/1-1 and by HGF under No. VH-NG-811. \\

\bigskip
Author Contributions:\\
CC prepared samples and performed measurements and quantified all experimental data.
AG, MG, WD, and CSF provided HAXPES beamtime and support.
CB provided thermodynamic analyses.
ML provided access and support to HR-TEM.
KR and ML contributed thermodynamic calculations of Eu silicides.
MM provided organization, discussions and beamtime support.\\
Competing financial interests:\\
The authors declare no competing financial interests.



\end{document}